\documentstyle[12pt]{article}
\newcommand{\lar}[1]{\langle #1 \rangle}

\newcommand{\bel}[1]{\begin{equation}\label{#1}}
\newcommand{\ee}{\end{equation}}

\begin{document}

\begin{center}
{\Large Accurate mapping of quantum Heisenberg magnetic models of spin $s$ on
strong-coupling magnon systems}\\[0.5cm]

{\large Bang-Gui Liu$^*$ and Gerd Czycholl\\}

Institute for Theoretical Physics, Universit\"at Bremen, D--28334 Bremen,
Germany\\[2cm]
\end{center}

An infinite-$U$ term is introduced into the Holstein-Primakoff-transformed
magnon hamiltonian of quantum Heisenberg magnetic models of spin $s$.
This term removes the unphysical spin
wave states on every site and truncates automatically the expansion in
powers of the magnon occupation operator. The resultant strong-coupling
magnon hamiltonians are accurately equivalent to the original spin
hamiltonians. The on-site $U$ levels and their implications are studied.
Within a simple decoupling approximation for our strong-coupling magnon models
we can easily reproduce the results for the (sublattice) magnetizations
obtained previously for the original spin model. But our bosonic
hamiltonians without any unphysical states allow for substantially improved
values for the spectral weight in the ground state and for lower
ground-state energies than those obtained within previous
approximations. \\

\vspace{3cm} \noindent PACS numbers: 75.10Jm and 75.30Ds

\newpage

\section{Introduction}

Quantum Heisenberg magnetic models, including ferromagnetic (FM) and
antiferromagnetic (AFM) ones, are well-accepted models for insulating
ferromagnets and antiferromagnets. The exact analytical solution is limited
to one dimension and FM ground states. For general parameters one has to
turn to some approximation methods or to numerical work. As for analytical
methods, one can work directly with the original Heisenberg model, i.e.
using spin operators and their algebra. In this category are decoupling
approaches\cite{bogo,pu}, the spherical approach\cite{sph},
the projection method\cite{ks}, and an isotropic
decoupling approach\cite{iswt}. The first one was a mean-field approximation
in the context of Green functions, the second one was for paramagnetic states,
the third one investigated the ground states only,
and the last one was for short-range magnetic correlation.
Anyway, a treatment in terms of the bosonic magnon operators should be
advantageous because of the simpler commutation rules and the Bose
statistics. For this purpose one has to map the Heisenberg model on a spin
wave model using the well known Holstein-Primakoff\cite{hp}
or the Dyson\cite{d}
transformation. In fact, many experimental physicists tend to describe their
experimental results in terms of spin wave theory\cite{exp,kampf}.
But if
one chooses the Holstein-Primakoff transformation, a series
expansion of a square root term in powers of magnon occupation operators is
necessary\cite{hp,hp1};
and one has to break the spin operator relation $(S^{+})^{\dagger
}=S^{-}$ if one chooses Dyson's spin wave transformation\cite{d,d1}.

The most simple spin wave theory is linear spin wave theory\cite{mattis,lsw}.
Some nonlinear effects, namely essentially the next terms in the
expansion of the squareroot, are taken into account in nonlinear spin wave
theories\cite {mattis,hp,d,nlsw,kubo}.
But the Hilbert space on which the spin wave (magnon)
operators are defined is much larger than the physical Hilbert space. For
spin $s$ the Hilbert space for a single site has the dimension $2s+1$. But
the Hilbert space on which the magnon operators operate is infinite
dimensional. In 3D ferromagnets there should be only few spin wave
excitations at low temperature. As for antiferromagnets, there should be a
substantial number of magnons even at zero temperature as the sublattice
magnetization is less than $s$. The effect of the unphysical magnon states
on the physical quantities becomes more serious with increasing temperature,
because then a thermal occupation of the unphysical states becomes possible.
In the paramagnetic (PM) phase, the unphysical states lead to serious
problems. Lindg\aa{}rd and Danielsen \cite{ld}proposed the
matching-of-matrix-elements (MME) method, in which operators with a
complicated algebra (like spin operators) are expressed in terms of Bose
operators so that not only the commutation rules but also certain matrix
elements of the original operators remain unchanged and the matrix
elements between unphysical states vanish. This MME-method can
be considered to be a generalization of the Holstein-Primakoff
transformation. Spin hamiltonians like the Heisenberg model are mapped
on an interacting Bose model containing on-site interaction terms, but
they operate on a Hilbert space, which
has still a much larger dimension than the physical Hilbert space.
In Mattis's book\cite{mattis} a similarity transformation is used
to eliminate partly the unphysical states in FM phase. In Takahashi's
modified spin wave theory, the total number of magnons is fixed in
the PM phase\cite{takahashi,lbgnl}. But on a single site, the dimension of
the magnon Hilbert space is still much larger than $2s+1$.
Friedberg, Lee, and Ren \cite{flr} introduced an on-site interaction
term into a lattice Bose Hamiltonian and proved the equivalence of this Bose
model with a spin-wave model which is equivalent to a
spin-$\frac12$ Heisenberg model; they applied
the theorem to the anisotropic
ferromagnetic Heisenberg model in a magnetic field\cite{flr}.

In this paper we shall introduce a large-$U$ term into the Holstein-Primakoff
transformed magnon hamiltonian of quantum Heisenberg magnetic models of
any spin $s$. In
the $U\rightarrow \infty$ limit this term rigorously removes the unphysical
magnon states on every site so that the magnon Hilbert space is mapped
accurately on the original spin state space. At the same time it truncates
automatically the high power terms of the magnon operators arising in the
series expansion. This approach has some connections with earlier
attempts \cite{ld,flr}, but we present a formulation being valid for
arbitrary spin $s$ and apply it to the calculation of physical
quantities like the order parameter.
We shall study the hierarchy of the on-site $U$ levels and
its implication on the spin physics. Within a simple decoupling
approximation, we can easily reproduce results for the FM magnetization and
AFM sublattice magnetization which were obtained previously only in theories
working with the original spin operators and were better than the
results of conventional spin wave theories.
Furthermore our bosonic hamiltonians without
unphysical states allow us to obtain improved values for the renormalization
of the spectral factor at zero temperature and lower ground-state energies
than those of the existing approximations.\\

\section{Bosonic hamiltonians without unphysical states}

Our ferromagnetic (FM) and antiferromagnetic (AFM) Heisenberg hamiltonians
are defined by 
\begin{equation}
H=\pm \sum_{\langle ij\rangle }J_{ij}(\frac
12(S_i^{+}S_j^{-}+S_i^{-}S_j^{+})+S_i^zS_j^z)  \label{Heis}
\end{equation}
where $J_{ij}$ is positive and the summation is over the nearest neighbor
sites. Here the negative sign corresponds to the ferromagnetic case and the
positive sign to the antiferromagnetic case. For the AFM case it is better
to make a $\pi $ spin rotation for the operators in one of the two
sublattices. The rotated AFM hamiltonian reads 
\begin{equation}
H=\sum_{\langle ij\rangle }J_{ij}(\frac
12(S_i^{+}S_j^{+}+S_i^{-}S_j^{-})-S_i^zS_j^z)  \label{AFM}
\end{equation}
We choose Holstein-Primakoff (HP) transformation to transform the spin
operators into the magnon operators. 
\begin{equation}
S_i^{-}=a_i^{\dagger }\sqrt{2s-n_i},~~~S_i^{+}=\sqrt{2s-n_i}%
a_i,~~~S_i^z=s-n_i;~~~~~n_i=a_i^{\dagger }a_i  \label{HP}
\end{equation}
The magnon operators $a_i$ are standard bosonic operators. We prefer HP
transformation because Dyson transformation breaks the conjugate relation of
the spin operators. When substituting the transformation (\ref{HP}) into the
hamiltonians, one obtains the following FM hamiltonian 
\begin{equation}
H=\sum_i\epsilon a_i^{\dagger }a_i-\sum_{\langle ij\rangle }J_{ij}[\frac
12(a_i^{\dagger }\sqrt{2s-n_i}\sqrt{2s-n_j}a_j+{\rm h.c.})+a_i^{\dagger
}a_ia_j^{\dagger }a_j]-\frac 14\epsilon N  \label{FMb}
\end{equation}
and the following AFM hamiltonian 
\begin{equation}
H=\sum_i\epsilon a_i^{\dagger }a_i+\sum_{\langle ij\rangle }J_{ij}[\frac
12(a_i^{\dagger }a_j^{\dagger }\sqrt{2s-n_i}\sqrt{2s-n_j}+{\rm h.c.}%
)-a_i^{\dagger }a_ia_j^{\dagger }a_j]-\frac 14\epsilon N  \label{AFMb}
\end{equation}
Here $N$ is the total number of the sites, $\epsilon =JZ/2$, and
$J$ is the exchange constant in the isotropic case or
the largest of the $J_{ij}$ in the anisotropic case and $Z$ is the
coordination number.
Since the operator $a_i$ is a standard Bose operator, it operates on an
infinite dimensional Hilbert space. But the physical Hilbert space
corresponding to a single site is spanned by only $2s+1$ states. The extra
states are unphysical. The magnon hamiltonians (\ref{FMb}) and (\ref{AFMb})
are not equivalent to the original spin hamiltonians (\ref{Heis})
and (\ref{AFM}) if the unphysical states are not removed. For
the exact FM ground state it is expected that there is no bosonic excitation
so that the unphysical states remain unoccupied at zero temperature. The
higher the temperature is, the more serious problems may arise, if the
thermal occupation of the unphysical states becomes possible. In the AFM
case even the ground state has some substantial bosonic excitations. The
largest discrepancy appears in the paramagnetic (PM) phase where the average
spin in conventional magnon theory becomes very unreasonable if one tries to
calculate it without introducing some constraints.

To remove completely the effect of the unphysical magnon states on every
site, we can make their energy levels infinitely higher than those of the
physical spin states. We introduce a large $U$ term into the
Holstein-Primakoff transformed hamiltonians. This $U$ term resembles the
strong coupling positive $U$ term in the Hubbard model of electronic
systems. But it is not dynamical, it is introduced only as a constraint to
raise the unphysical states infinitely high in energy from the physical
states. This means that our new hamiltonians $H^{\prime }$ are composed of
the original hamiltonians $H$ and the following $U$ terms. 
\[
H_U=\frac 1{(2s+1)!}Ua_i^{\dagger (2s+1)}a_i^{(2s+1)} 
\]
In fact, $U$ should be considered to be infinite. Therefore, the unphysical
states should be pushed infinitely high in energy from the physical states
by the $U$ term. For the half spin case, the resultant FM hamiltonian reads 
\begin{equation}
H^{\prime }=\sum_i(\epsilon a_i^{\dagger }a_i+\frac 12Ua_i^{\dagger
2}a_i^2)-\sum_{\langle ij\rangle }J_{ij}[\frac 12(a_i^{\dagger
}a_j+a_ia_j^{\dagger })+a_i^{\dagger }a_ia_j^{\dagger }a_j]-\frac 14\epsilon
N  \label{FMham}
\end{equation}
and the AFM hamiltonian reads
\begin{equation}
H^{\prime }=\sum_i(\epsilon a_i^{\dagger }a_i+\frac 12Ua_i^{\dagger
2}a_i^2)+\sum_{\langle ij\rangle }J_{ij}[\frac 12(a_i^{\dagger }a_j^{\dagger
}+a_ia_j)-a_i^{\dagger }a_ia_j^{\dagger }a_j]-\frac 14\epsilon N
\label{AFMham}
\end{equation}
 Here we need no chemical potential
since the unphysical magnon states have been removed rigorously by the
infinite-$U$ term. Furthermore the square root terms $\sqrt{2s-a^{\dagger }a%
}$ in the hamiltonians (\ref{FMb}), (\ref{AFMb}) have been expanded
as
\begin{equation}
\begin{array}{ll}
\sqrt{2s-a^{\dagger }a}= & \displaystyle \sqrt{2s}-(\sqrt{2s}-\sqrt{2s-1}%
)a^{\dagger }a+(\sqrt{2s}-2\sqrt{2s-1}+\sqrt{2s-2})\frac{a^{\dagger 2}a^2}{2!%
} \\ 
& \displaystyle +({\rm terms\;of\;}a^{\dagger n}a^n,\;n\geq 3)
\end{array}
\label{root}
\end{equation}
and all terms  $%
a^{\dagger n}a^n$ with $n\geq 2s$ can be neglected after introduction of
the U-terms because their energy is already shifted to infinity by the
U-term.
In the cases of $s=\frac 12,$ all operator product terms including $%
a_j^{\dagger n}a_i^m$ ($m>1$ and/or $n>1$) disappear automatically so that
the hamiltonians (\ref{FMham}) and (\ref{AFMham}) are very simple. For
larger spin $s$, there are more terms resulting from the expansion of the
square root because the terms including $a_i^{\dagger n}a_j^m$ ($m\leq 2s$
and $n\leq 2s$) are allowed. This expansion is different from the $1/s$
expansion in the spin wave theories. We make no approximation in the
expansion (\ref{root}). The $U$ term not only pushes the unphysical states
infinitely high from the physical states but also truncates automatically
the expansion of the square root. Our expansions of $S^+_i$ and $S^-_i$
are composed of only $2s$ terms and $S^z_i$ is $s-a^{\dagger}_ia_i$,
being different from the infinite sums of the three spin operators in
Ref \cite{ld}. Our mapping works for ferromagnetic and antiferromagnetic
Heisenberg models of any spin $s$, being in contrast to the equivalence
theorem which works only for half-spin system\cite{flr}.\\

\section{On-site $U$ levels}

To study the effect of the $U$ term, we first study the on-site $U$ levels.
In the half-spin case, we have the following commutation relations: 
\begin{equation}
\lbrack a^{\dagger 2}a^2,a^{\dagger p}a^q]_{-}=[p(p-1)-q(q-1)]a^{\dagger
p}a^q+2(p-q)a^{\dagger (p+1)}a^{(q+1)}  \label{com}
\end{equation}
For $q=0$ and $p=n$, we obtain by application of this expression on the
magnon vacuum: 
\begin{equation}
H_U^{1/2}|n\rangle =\frac{n(n-1)U}2|n\rangle  \label{state1}
\end{equation}
where $H_U^{1/2}=\frac U2a^{\dagger 2}a^2$ and $|n\rangle =a^{\dagger
n}|0\rangle $. $|n\rangle $ is an eigenstate of $H_U^{1/2}$. The magnon
states for the lowest five $U$ levels are shown in the left part of Figure
1. The magnon vacuum state $|0\rangle $ and the single magnon state $%
a_i^{\dagger }|0\rangle $ correspond to the two physical spin states. The
multiple magnon states $a_i^{\dagger n}|0\rangle $ $(n>1)$ are separated
from these physical states by an energy of the magnitude of $U$ and are thus
projected out from the Hilbert space in the limit $U\rightarrow \infty $.
Therefore, we expect $\langle a_i^{\dagger n}a_i^n\rangle =0$ $(n\geq 2)$
when $U$ tends to $\infty$.\\

As for the case of spin $1$, we have the following operator equality. 
\begin{equation}
\begin{array}{c}
\lbrack a^{\dagger 3}a^3,a^{\dagger p}a^q]_{-}=\displaystyle %
[p(p-1)(p-2)-q(q-1)(q-2)]a^{\dagger p}a^q+ \\ 
~\displaystyle 3[p(p-1)-q(q-1)]a^{\dagger (p+1)}a^{(q+1)}+3(p-q)a^{\dagger
(p+2)}a^{(q+2)}
\end{array}
\label{com1}
\end{equation}
The on-site $U$ part of the hamiltonian is $H_U^1=\frac U6a^{\dagger 3}a^3$.
We obtain the following eigenstate equation. 
\begin{equation}
H_U^1|n\rangle =\frac{n(n-1)(n-2)U}6|n\rangle  \label{state2}
\end{equation}
The states within the first five on-site levels are shown in the right part
of Figure 1. At the ground level are the magnon vacuum $|0\rangle $, single
magnon state $a_i^{\dagger }|0\rangle $, and double magnon state $%
a_i^{\dagger 2}|0\rangle $. They correspond to the three physical states of
the spin operator: $-1,0,1$. Other states are separated from the physical
states by energies of at least $U$. Therefore, we expect $\langle
a_i^{\dagger n}a_i^n\rangle =0$ $(n\geq 3)$ when $U$ tends to $\infty$.\\

For higher spins $s$, we can derive some operator equations similar to (\ref
{com}) and (\ref{com1}). Always we have $2s+1$ magnon states on the ground
level, corresponding to the total physical spin states, and the unphysical
states are separated from the physical states by energies of order $U$ . We
expect $\langle a_i^{\dagger n}a_i^n\rangle =0$ $(n\geq 2s+1)$ when $U$
tends to $\infty$.\\

\section{First-order decoupling approximation}

Since the $U$ is very large, we cannot apply Hartree-Fock approximation to
the magnon hamiltonians. Now we study the FM and AFM systems of half-spin in
a first-level decoupling approach of its equation of motion. We choose $a_i$
as our dynamical variable.

{\em The FM case: }Using the Zubarev notation for the commutator Green
function
\begin{equation}
\langle \langle A;B\rangle \rangle _z=-i\int_{-\infty }^{+\infty }\theta
(t)\langle [A(t),B]\rangle e^{izt}
\end{equation}
we get the equation of motion
\begin{equation}
(z-\epsilon )\langle \langle a_i;a_j^{\dagger }\rangle \rangle _z=\delta
_{ij}-\frac 12\sum_lJ_{il}\langle \langle a_l;a_j^{\dagger }\rangle \rangle
_z-\sum_lJ_{il}\langle \langle a_in_l;a_j^{\dagger }\rangle \rangle
_z+U\langle \langle a_i^{\dagger }a_i^2;a_j^{\dagger }]\rangle \rangle _z
\label{e1}
\end{equation}
In the above equation the third term comes from the inter-site interaction
and the fourth term comes from the on-site large $U$ interaction. Without
these two terms, one recovers the conventional FM linear spin wave theory.
Since $U$ is very large, the fourth term cannot be neglected. We have to
write down its equation of motion to get closed equations.
\begin{equation}
(z-\epsilon -U-JZn)\langle \langle a_i^{\dagger }a_i^2;a_j^{\dagger }\rangle
\rangle _z=2n\delta _{ij}-n\sum_lJ_{il}\langle \langle a_l;a_j^{\dagger
}\rangle \rangle _z  \label{e2}
\end{equation}
Here $n_l=a_l^{\dagger }a_l$ denotes the magnon occupation operator, $%
n=\langle n_l\rangle $ its thermal expectation value. In the latter equation
other higher order Green functions have been neglected because they involve
an even higher multiple magnon occupation of a single site and, therefore,
vanish for infinite $U$. Furthermore a decoupling approximation between
magnon and occupation operators for different sites has been made, but
operators operating on the same site are not decoupled. Since $U$ tends to
infinite and $z$ is of order of $J$, the prefactor on the left hand side can
be simplified into $-U$. Substituting it into (\ref{e1}) and using once more
the mentioned decoupling approximation, we get 
\begin{equation}
(z-(1-2n)\epsilon )\langle \langle a_i;a_j^{\dagger }\rangle \rangle
_z=(1-2n)\delta _{ij}-\frac 12(1-2n)\sum_lJ_{il}\langle \langle
a_i;a_j^{\dagger }\rangle \rangle _z  \label{e3}
\end{equation}
By Fourier transformation we obtain for the $\vec k$-dependent Green function
\begin{equation}
G_{\vec k}(z)=\frac 1N\sum_{i,j}\langle \langle a_i;a_j^{\dagger }\rangle
\rangle _ze^{i\vec k(\vec R_i-\vec R_j)}=\displaystyle \frac{1-2n}{%
z-(1-2n)\epsilon (1-r_k)}  \label{e4}
\end{equation}
where 
\begin{equation}
r_k=\frac 1Z\sum_{\vec \Delta }e^{i\vec k\vec \Delta }
\end{equation}
denotes the (dimensionless) magnon dispersion characteristic for the lattice
under consideration; $\vec \Delta $ denotes the nearest neighbor vectors.
For $d$-dimensional simple cubic lattice, we have
$$r_k=\frac 1d\sum_{i=1}^d\cos k_i$$
The self-consistency equation to determine $n$ reads 
\begin{equation}
n=\displaystyle (1-2n)\frac 1N\sum_{\vec k}1/[\exp \frac{JZ(1-2n)(1-r_k)}{2T}%
-1]  \label{e5}
\end{equation}
This equation is equivalent to that expressed in terms of spin operator average,
$\lar{S^z_i}$, obtained by Bogoliubov\cite{bogo}. In
the limit $T\rightarrow 0$ we obtain $n=0$ or $1$ as solutions of the above
equation. These two solutions correspond to $\langle S_i^z\rangle =\frac 12$
or $-\frac 12$, respectively , since $S_i^z=\frac 12-n_i.$ When temperature
increases to the Curie Temperature $T_c$ , the two branches converge to $%
n=\frac 12,$ or $\langle S_i^z\rangle =0$ . For finite but small $T$ we
obtain $\langle S_i^z\rangle =\frac 12-\alpha (T/J)^{3/2}$ where $\alpha $
is a positive constant. When $T\rightarrow T_c$ , $(1-2n)$ tends to zero so
that we can derive an asymptotic expression of $\langle S_i^z\rangle $ as
follows. 
\begin{equation}
\langle S_i^z\rangle \propto \sqrt{1-T/T_c},~~~~~T_c=JZ/4P,~~~~~P=\frac
1N\sum_{\vec k}1/(1-r_k)  \label{asymp}
\end{equation}
In two or one dimensions the $\vec k$-integration diverges, i.e. $P=\infty $
so that $T_c=0$ in accordance with the Mermin-Wagner theorem \cite{mw}.

For $T\leq T_c$ in three dimensions, we obtain two solutions. These two
solutions correspond to the two degenerate ferromagnetic solutions . The
resulting order parameter, i.e. the magnetization is shown in Figure 2. The
3D results of the conventional nonlinear and linear spin wave theories are
also presented for comparison\cite{mattis}.
The nonlinear spin wave theory produces an unphysical  first-order
transition at $T=0.98J$\cite{mattis}. The
magnetization obtained within our strong-coupling magnon theory
according to (\ref{e5}) is obviously an essential improvement in the whole
temperature regime; we obtain as the critical (Curie) temperature $T_c =
0.989J$ whereas the series expansion result for $T_c$ is
$T_c=0.889J$\cite{jc}. It is interesting to compare this $T_c$ result to the
spherical approximation for paramagnetic phase\cite{sph}.
The spherical approximation result, $T_c=0.82J$, is slightly smaller than
the numerical results, whereas our result $T_c=0.989J$
is on the higher temperature side of the numerical results.
Our ground state energy is $-0.25Jd$ per site in two and three
dimensions, as it should be. \\

{\em The AFM case: }In this case the Green function $G_{ij}(z)=\langle
\langle a_i;a_j^{\dagger }\rangle \rangle _z$ only is not sufficient but we
have to consider also the ''anomalous'' one-particle Green function $%
F_{ij}(z)=\langle \langle a_i^{\dagger };a_j^{\dagger }\rangle \rangle _z$ .
We obtain the following equation of motion for the Green function $G_{ij}(z)$.
\begin{equation}
(z-\epsilon )G_{ij}=\delta _{ij}+\frac
12\sum_lJ_{il}F_{lj}-\sum_lJ_{il}\langle \langle a_in_l;a_j^{\dagger
}\rangle \rangle _z+U\langle \langle a_i^{\dagger }a_i^2;a_j^{\dagger
}\rangle \rangle _z  \label{e11}
\end{equation}
In the above equation the third term comes from the inter-site interaction
and the fourth term comes from the on-site large $U$ interaction. Without
these two terms, one obtains the existing AFM linear spin wave theory. In
the same way as in the FM case, we have to write down a further equation of
motion for the $U$ term. It reads 
\begin{equation}
(z-\epsilon -U-JZn)\langle \langle a_i^{\dagger }a_i^2;a_j^{\dagger }\rangle
\rangle _z=2n\delta _{ij}+n\sum_lJ_{il}F_{lj}(z)  \label{e12}
\end{equation}
where we have again decoupled magnon occupation number operators and single
magnon operators at different lattice sites. Inserting this into (\ref{e1})
and using the decoupling approximation once more, we get 
\begin{equation}
(z-(1-2n)\epsilon )G_{ij}=(1-2n)\delta _{ij}+\frac 12(1-2n)\sum_lJ_{il}F_{lj}
\label{e13}
\end{equation}
In the same way, we obtain: 
\begin{equation}
(z-(1-2n)\epsilon )F_{ij}=-\frac 12(1-2n)\sum_lJ_{il}G_{lj}  \label{e14}
\end{equation}
These two equations are closed and yield the Green functions. This
approximation is similar to the first-order Hubbard approximation of the
electronic Hubbard model. Since $S_i^z=1/2-n_i$, it is clear that the above
result is equivalent to results obtained previously for the spin model \cite
{bogo,pu}. The Green functions are given by 
\begin{equation}
\begin{array}{l}
G_k=(1-2n)[z+(1-2n)\epsilon ]/[z^2-\epsilon ^2(1-2n)^2(1-r_k^2)] \\ 
F_k=-(1-2n)^2\epsilon r_k/[z^2-\epsilon ^2(1-2n)^2(1-r_k^2)]
\end{array}
\label{gfafm}
\end{equation}
The self-consistent equation to determine the $n$ is given by 
\begin{equation}
\frac 12=(\frac 12-n)\frac 1N\sum_k\frac 1{\sqrt{1-r_k^2}}\coth \frac{\omega
_k}{2T}  \label{e15}
\end{equation}
where $\omega _k=(\frac 12-n)JZ\sqrt{1-r_k^2}$. In three dimensions we get
for the zero-temperature sublattice magnetization, $S_0=0.4325$, and the N\'eel
temperature is:$T_N=0.989J$. The sublattice magnetization
as a function of $T$ is
shown in Figure $3$. The N\'eel temperature is better than that of the
conventional spin
wave theories because the high temperature expansion result is $0.951J$\cite
{jc}. For small temperature $T$ we obtain $\langle S^z\rangle =S_0-\eta
(T/J)^2$ . When temperature tends to $T_N$, the sublattice magnetization
has the following asymptotic expression.
\begin{equation}
\langle S^z\rangle \propto \sqrt{1-T/T_N},~~~~~T_N=JZ/4P  \label{asymp1}
\end{equation}
The sublattice magnetization as a function of temperature is shown in Figure 3.
The results of the linear spin wave theory and a nonlinear spin wave theory are
also presented for comparison.
The nonlinear theories produce
an unphysical first order transition at $T=1.11J$ in three
dimensions\cite{tl}, being similar to the FM case.
The result from (\ref{e15})
is best in the whole temperature region. In two dimensions $T_N=0$ and the
zero-temperature sublattice magnetization
is equivalent to $0.3587$ , being larger
than the results of the spin wave theories and a series expansion result\cite
{singh} in which the spin wave behavior was used in their extrapolation. But
it is consistent to a Monte Carlo result $0.34\pm 0.01$\cite{carl}. A Green
function Monte Carlo result is $0.31\pm 0.02$\cite{tc}. In one dimension
the average sublattice magnetization is zero even at zero temperature.
This is
consistent with the Mermin-Wagner theorem\cite{mw}. It is clear that the
inter-site coupling modifies the spectra and the $U$ term reduces the
spectral weight. Without these two terms, we should get the linear spin wave
theory. But the $U$ term is necessary to remove the unphysical states. The
linear spin wave theory overestimated the spectral weight by about thirty
percent in terms of a study on the sum rules and spin excitations of the
quantum AFM Heisenberg models\cite{str}.

The $U$ term does not contribute to the ground state energy. Our ground
state energies in two and three dimensions are  $%
E_0^{AFM}/\epsilon N=-0.327$ and $-0.297$, respectively. The ground state
energies are a little higher than the existing results of spin wave
theories\cite{lsw,nlsw,takahashi,lbgnl}. On the other hand, the spectral
factor $f=1-2n$ is less than $1$. Therefore, some improvement to the ground
state is desirable.\\ 

{\em The AFM ground states in a relaxed decoupling: } To improve our
approximation, we relax the constraint of the decoupling by permitting the
decoupling of the operators on a site without changing the position of the
operator product in the on-site $U$ level hierarchy. In this approximation,
the inter-site correlation functions enter the spectral renormalization
factor $f$ so that we obtain the following nonlinear equation set of two
variables at zero temperature. 
\begin{equation}
\begin{array}{rcl}
\displaystyle \frac 12 & = & \displaystyle (\frac 12-n)\frac 1N\displaystyle %
\sum_k\frac 1{\sqrt{1-r_k^2}} \\ 
\xi  & = & \displaystyle -(\frac 12-n)\frac 1N\sum_k\frac{r_k^2}{\sqrt{%
1-r_k^2}}
\end{array}
\label{ee1}
\end{equation}
Now our magnon spectrum is defined by $\omega _k=\frac 12JZf\sqrt{1-r_k^2}$
and our spectral factor is defined by $f=1-2n-2\xi $. For the ground state
we obtain the same $n$ and $\xi $ as the above. Therefore we obtain $%
(E_0^{AFM}/\epsilon N,S_0^z,f)=(-0.3106,0.4325,1.084)$ in three dimensions
and $(-0.345,0.3587,1.113)$ in two dimensions. The renormalization factors
are acceptable and the ground state energies are lower than those of the
spin wave theories\cite{lsw,nlsw,lbgnl,hp1,d1} and others available\cite
{carl,tc,singh,ks,pu}. The details are summarized in Table 1. The ground
state energies, $E_0$, in Table 1 are given in units of $JdN$ where $d$ is
the dimension.\\ 

\section{Discussion and summary}

We have studied the half-spin strong-coupling magnon hamiltonians without
any unphysical states in a simple Hubbard-like decoupling approximation. For
higher spins the strong-coupling hamiltonians can be treated similarly.
Following the routine described by Fulde\cite{fulde}, we can also treat our
magnon hamiltonians by means of the projection method. In the above simple
decoupling approximation we obtain the same sublattice spins as that
obtained directly from the original spin hamiltonians, but our spectral
renormalization factors at zero temperature are improved substantially and
our ground state energies are lower than those of existing
approximations. From Figure 2 and Figure 3 it is clear that
our strong-coupling magnon hamiltonians in the
simple decoupling approximation improve the conventional spin wave theories.
The nonlinear spin wave
theories produce unphysical first-order transitions. There have been many
versions for the nonlinear spin wave theory, but the main feature and
drawback is similar in all these versions. But our strong-coupling magnon
theories do not lead to such unphysical behavior,
because all unphysical states in
the Hilbert space have been removed, and is, therefore, advantageous over
the original spin model and the conventional magnon hamiltonian.
Compared to Ref.\cite{flr}, where  the introduction of a similar
strong-coupling U-term was suggested, we presented here a formulation,
which works for
ferro- and antiferromagnetic Heisenberg models of any spin $s$, and we
applied for the first time a Green function decoupling approximation to
this model and could calculate quantities like the order parameter
(magnetization), the critical temperature $T_c$, etc.\\

In summary, we introduce an infinite-$U$ term into the
Holstein-Primakoff magnon
hamiltonian of quantum Heisenberg magnetic models of any spin $s$.
This term rigorously
removes the unphysical magnon states on every site and at the same time
automatically truncates the expansion of the square root $\sqrt{1-n_i/s}$.
The resultant magnon
hamiltonians are accurately equivalent to the original spin hamiltonians. We
have studied the on-site $U$ levels and their implication on the spin
physics. Within a simple decoupling approximation we obtain physically
reasonable results for the FM magnetization and AFM sublattice magnetization
in agreement with existing results obtained for the original spin model. But
we obtain lower ground-state energies than those of the previous theories
because our hamiltonians are composed of the bosonic magnon operators and
free of unphysical states. \\ 

\section*{Acknowledgment}

BGL appreciately acknowledges financial supports from the Alexander von
Humboldt foundation (Jean-Paul-Str. 12, 53173 Bonn, Germany).
The authors are grateful to the referees for supplying Refs \cite{sph,ld,flr}.

\newpage
{\noindent{\large {\bf Table and Figure Captions}}}

\begin{description}
\item[Table 1: ]  The AFM ground state energy and sublattice spin available
in various approximations of quantum Heisenberg models of half-spin. The $E_0
$'s are in unit of $Jd$. LSW: the linear spin wave theory; NLSW: the
nonlinear spin wave theory; Series+SW: the series expansion method in which
some spin wave behavior was used in their extrapolation; MC: the Monte
Carlo; GFMC: Green function Monte Carlo; Projection: projection method by
spin operators; SGFMF: spin Green function mean-field; This work: The result
of this paper.\\[1.2cm]

\item[Fig 1: ]  {\it left: } On-site $U$ levels in the case of the half
spin. At the ground level there are only $|0\rangle$ and 
$a_i^{\dagger }|0\rangle$. {\it %
right: } On-site $U$ levels in the case of the unit spin. At the ground
level there are only $|0\rangle$, $a_i^{\dagger }|0\rangle$, 
and $a_i^{\dagger 2}|0\rangle$.

\item[Fig 2: ]  3D magnetization of the FM model
as function of temperature. The
solid line is for  the approximation in this paper; the dashed line
for nonlinear spin wave theories; the dotted line for linear spin wave
theory. The $T_c$'s are $0.989J$, $0.98J$, and $1.71J$, respectively. But a
high temperature expansion result is $0.889J$. The transition for the
nonlinear spin wave theory is of first order.

\item[Fig 3: ]   3D sublattice magnetization of the AFM model
as function of temperature. The
solid line is for  the approximation in this paper; the dashed line
for nonlinear spin wave theories; the dotted line for linear spin wave
theory. The $T_N$'s are $0.989J$, $1.107J$, and $1.70J$, respectively. But a
high temperature expansion result is $0.951J$. The transition for the
nonlinear spin wave theories is of first order.
\end{description}

\newpage

\begin{center}
\begin{tabular}{||c|c|c|c|c||}
\hline\hline
Approximation & 2D $S_0^3$ & 2D $E_0$ & 3D $S_0^3$ & 3D $E_0$ \\ \hline
LSW\cite{lsw} & 0.303 & -0.329 & 0.422 & -0.2985 \\ \hline
NLSW$^1\cite{hp1,d1}$ & 0.3069 &  &  &  \\ \hline
NLSW$^2$\cite{lbgnl,tl} & 0.303 & -0.335 & 0.422 & -0.301 \\ \hline
Series+SW\cite{singh} & 0.3025 & -0.3348 &  &  \\ \hline
MC\cite{carl} & 0.34$\pm 0.01$ & -0.335 &  &  \\ \hline
GFMC\cite{tc} & 0.31$\pm 0.02$ & -0.3346 &  &  \\ \hline
Projection\cite{ks} & 0.359 & -0.132 &  &  \\ \hline
SGFMF\cite{pu} & 0.3587 & -0.327 & 0.4325 & -0.297 \\ \hline
This work & 0.3587 & -0.327 & 0.4325 & -0.297 \\ \hline
This work (improved) & 0.3587 & -0.365 & 0.4325 & -0.309 \\ \hline\hline
\end{tabular}
\end{center}

\newpage
\begin{figure}[tp]
\setlength{\unitlength}{2pt} 
\begin{picture}(60,100)(0,0)
\put(30,0){\line(1,0){10}}
\put(30,20){\line(1,0){10}}
\put(30,40){\line(1,0){10}}
\put(30,60){\line(1,0){10}}
\put(30,80){\line(1,0){10}}
\put(35,0){\line(0,1){100}}

\put(15,0){\makebox(14,8)[r]{$0$}}
\put(15,20){\makebox(14,8)[r]{$U$}}
\put(15,40){\makebox(14,8)[r]{$2U$}}
\put(15,60){\makebox(14,8)[r]{$3U$}}
\put(15,80){\makebox(14,8)[r]{$4U$}}

\put(41,0){\makebox(50,8)[l]{$|0\rangle,a^{\dagger}|0\rangle$}}
\put(41,20){\makebox(50,8)[l]{$a^{\dagger 2}|0\rangle$}}
\put(41,40){\makebox(50,8)[l]{}}
\put(41,60){\makebox(50,8)[l]{$a^{\dagger 3}|0\rangle$}}
\put(41,80){\makebox(50,8)[l]{}}
\end{picture}
\par
\begin{picture}(60,100)(-65,-100)
\put(30,0){\line(1,0){10}}
\put(30,20){\line(1,0){10}}
\put(30,40){\line(1,0){10}}
\put(30,60){\line(1,0){10}}
\put(30,80){\line(1,0){10}}
\put(35,0){\line(0,1){100}}

\put(15,0){\makebox(14,8)[r]{$0$}}
\put(15,20){\makebox(14,8)[r]{$U$}}
\put(15,40){\makebox(14,8)[r]{$2U$}}
\put(15,60){\makebox(14,8)[r]{$3U$}}
\put(15,80){\makebox(14,8)[r]{$4U$}}

\put(41,0){\makebox(50,8)[l]{$|0\rangle,a^{\dagger}|0\rangle, 
a^{\dagger 2}|0\rangle$}}
\put(41,20){\makebox(50,8)[l]{$a^{\dagger 3}|0\rangle$}}
\put(41,40){\makebox(50,8)[l]{}}
\put(41,60){\makebox(50,8)[l]{}}
\put(41,80){\makebox(50,8)[l]{$a^{\dagger 4}|0\rangle$}}
\end{picture}
\label{fig1}
\end{figure}


\begin{thebibliography}{*}
\bibitem{tl} See \cite{takahashi,lbgnl}. But those two papers were on two
dimensional and quasi-two dimensional cases. Extension of those to three
dimensions is made for comparison with the current result
\end{thebibliography}
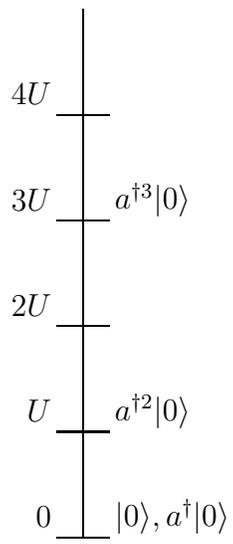
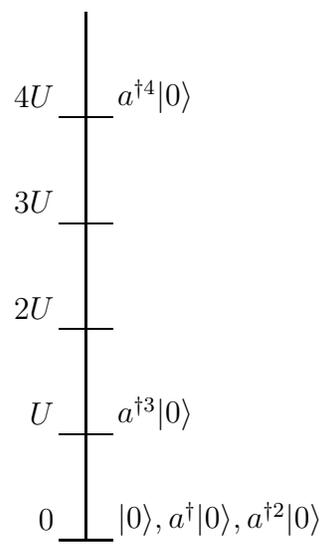
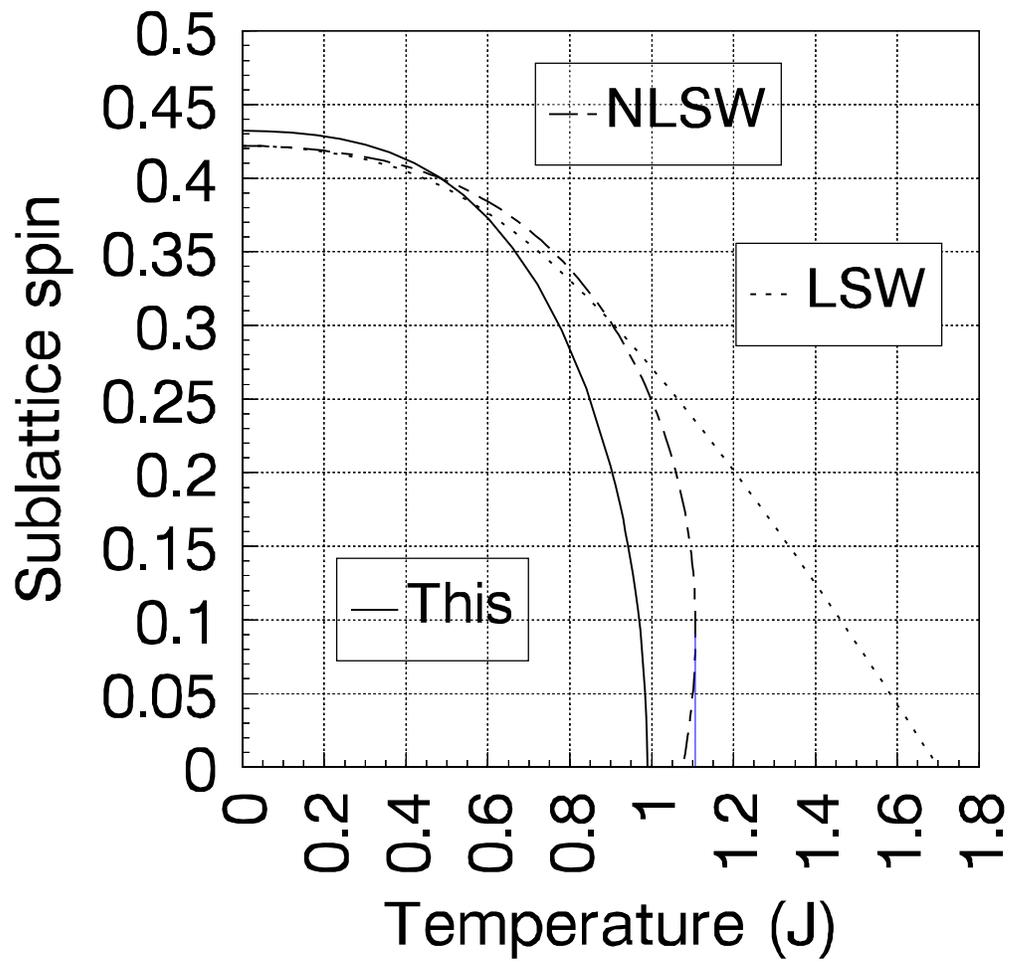
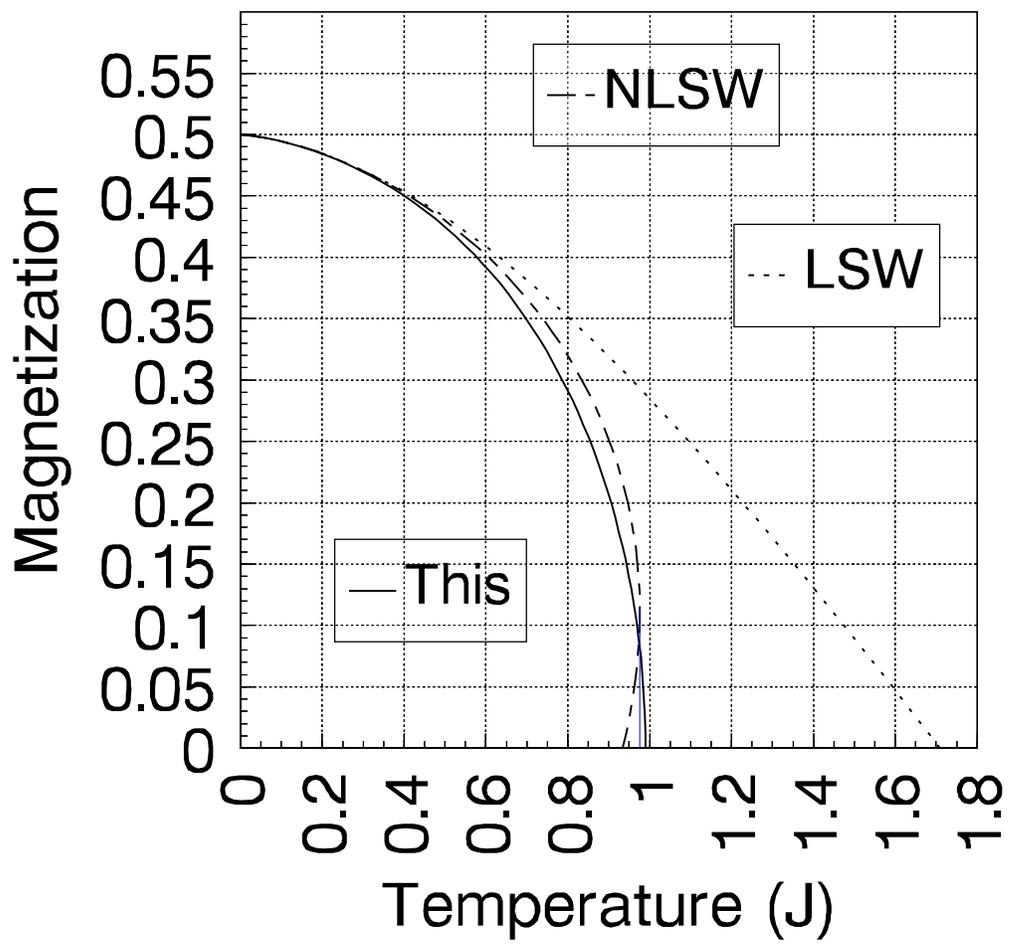
\end{document}